\documentclass[11pt]{revtex4-1}

\usepackage{graphicx}

\begin{document}

\title{Dynamical System Analysis for DBI Dark Energy interacting with Dark Matter}
\author{Nilanjana Mahata\footnote {nilanjana\_mahata@yahoo.com}}
\author{Subenoy Chakraborty\footnote {schakraborty.math@gmail.com}}

\affiliation{Department of Mathematics, Jadavpur University, Kolkata-700032, West Bengal, India.}



\begin{abstract}
 A dynamical system analysis related to Dirac-Born-Infeld (DBI) cosmological model has been investigated in this present work.  For  spatially flat FRW space time, the Einstein  field equations  for  DBI scenario  has been used to study the dynamics of DBI dark energy interacting with dark matter. The DBI dark energy model is considered as a scalar field with a non standard kinetic energy term. An interaction between the DBI dark energy  and dark matter is considered  through a phenomenological interaction between DBI  scalar field and the dark matter fluid.  The field equations are reduced to an autonomous dynamical system by a suitable redefinition of the basic variables.  The potential of the DBI scalar field  is assumed to be exponential. Finally, critical points are determined, their nature have been analyzed and corresponding cosmological scenario has been discussed.
\end{abstract}

\maketitle



\section{Introduction}
 Recent cosmological observations of Type Ia Supernovae  strongly indicate that the universe at present  has an accelerated expansion [1, 2 ]. This has been supported by subsequent observations  from Cosmic Microwave Background Radiation [CMBR][3, 4], Baryon Accoustic Oscillation [5] etc. In general, within the framework of Einstein gravity, this late time acceleration is attributed to  \textit{dark energy} (DE) having negative pressure .  Though the nature of dark energy is still unknown,  the simplest choice for dark energy is \textit{cosmological constant} or vacuum energy density which fits well for wide range of astronomical data. But  \textit{fine tuning }and \textit{coincidence problem}  are significant problems associated with cosmological constant [6, 7, 8]. To alleviate these problems, scalar fields having variable equation of state are introduced . Various  scalar field models of dynamical DE like quintessence [9,10,11], K-essence [12,13], Phantom [14, 15], tachyon [16, 17],  dilatonic ghost condensate [18], quintom [19, 20],  etc have been investigated [21] . \\

 On the other hand, from string-theoretic point of view the early accelerated expansion (i.e inflation) can be described by Dirac-Born-Infeld (DBI) inflation [22, 23, 24]. This model is  a special case of K- inflation models [25\emph{}] and is characterized by the open string sector through dynamical Dp-branes. It is found that the simplest DBI models are effectively indistinguishable from the usual (field theoretic) slow-roll models of inflation .
 In the present work , we shall examine whether the DBI model can explain the observed late time acceleration, choosing the DBI scalar field as DE. As the density of dark matter is comparable to dark energy in the present universe, so it is reasonable to consider an interaction between the two dark components.  The evolution equations are converted into an autonomous system by suitable transformation of the basic variables and  a phase space analysis is done. Finally, we check for any late-time attaractor solution in the phase space. The plan of the paper is as follows : Section II deals with the basic equations related to DBI model. Autonomous system  has been  constructed in section III and analysis of crtical points is presented in section IV.
\section{ Basic Equations}
In a four-dimensional spatially flat Friedman-Robertson-Walker (FRW) spacetime filled with a non-cannonical scalar field of type DBI , the energy density and pressure of DBI scalar field are given by
 \begin{equation}
 \rho_{\phi} = \frac{\nu^{2}}{\nu + 1}\dot{\phi}^{2}+ V(\phi)
 \end{equation}

 and
 \begin{equation}
  p_{\phi} =  \frac{\nu}{\nu + 1}\dot{\phi}^{2} - V(\phi)
\end{equation}
where $\nu $ has the form of a Lorentz boost factor,
\begin{equation}
  \nu =  \frac{1}{\sqrt{1- f(\phi)\dot{\phi}^{2}}},
\end{equation}
 $ V(\phi) > 0 $ is the potential, $ f(\phi) > 0  $ is the warp factor and $ \sqrt{f(\phi)}\dot{\phi}$ may be interpreted as proper velocity of the brane.  We assume the customary barotropic equation of state for the dark fluid of the form $ p = (\Gamma - 1) \rho $, where $ \Gamma  = \frac{\nu\dot{\phi^{2}}}{\rho}$ is the barotropic index of the DBI field. Also, positivity of the potential restricts $ \Gamma $ as $ 0 ~~ \leq~~ \Gamma ~~\leq ~~\frac{\nu + 1}{\nu}  $.\\
 In recent past, Guendelman et al [26] have constructed a unified model of DE and dark matter (DM) using a gravitating scalar field having similar non-conventional kinetic term and finds equivalent  effects. Further , the scalar field has a non standard DBI like Lagrangian  density and it corresponds to tachyonic scalar field by suitable restrictions on the potential.\\
  Now considering a flat FRW metric with scale factor a(t), the field equations are
  \begin{equation}
   \dot{H}  = - \frac{1}{2}[\nu\dot{\phi^{}} + \rho_{m}( 1 + \omega_{m})]
\end{equation}
and
\begin{equation}
 3H^{2} = \rho_{m} + \rho_{\phi},
\end{equation}
where $ \rho_{\phi}$ and $\rho_{m}$ are the energy densities of dark energy and dark matter respectively and  \\$ H = \frac{\dot{a}}{a} $ is Hubble rate of expansion. The
Density parameters are defined as
$ \Omega_{\phi} = \frac{\rho_{\phi}}{3H^{2}}$,  $ \Omega_{m} = \frac{\rho_{m}}{3H^{2}}$ with the condition $ \Omega_{\phi} + \Omega_{m} = 1 $. At present our universe is largely dominated by dark matter and dark energy whereas all types of other matters ( i.e baryonic) are insignificant. Further dark energy has a repulsive effect while other matters are attractive. So the interaction between them is considered as weak. However, interaction models are favoured by observed data obtained from the Cosmic Microwave Background (CMB) [27] and matter distribution at large scales [28].\\
 Hence, if the rate of creation/ annihilation between DBI scalar field and dark matter fluid be Q, we have the conservation equation as
\begin{equation}
\dot{\rho_{m}}+ 3H ( 1 + \omega_{m})\rho_{m} = Q
\end{equation}
and
\begin{equation}
\dot{\rho_{\phi}}+ 3H( 1 + \omega_{\phi})\rho_{\phi} = - Q
\end{equation}
where dot denotes differentiation with respect to cosmological time,  $ \omega_{\phi},~ \omega_{m} $ are corresponding equation of state parameters for dark energy and dark matter respectively.

The generic nature of the interaction $ Q > 0 $ indicates a flow of energy from DE to DM  while $ Q < 0 $ implies the reverse and is not permissible for the validity of second law of thermodynamics [29]. As there exists no fundamental theory which specifies coupling between dark energy and dark matter, so our coupling models will necessarily be phenomenological. However, one can make a comparative study between different coupling terms from physical or other natural ways. In the present work we choose the coupling term as a linear combination of the two energy densities i.e
\begin{equation}
Q = 3H (\alpha_{m}\rho_{m} + \alpha_{\phi}\rho_{\phi})
\end{equation}
$ \alpha_{m}$ and $ \alpha_{\phi}$ are dimensionless coupling parameters ( such that $ \mid \alpha_{m}\mid \ll 1 $ , $ \mid \alpha_{\phi}\mid \ll 1 $ ). H is introduced on dimensional ground and the factor 3 is due to mathematical convenience.\\

From (1), (4), (7) and (8) the evolution of the DBI scalar field takes the form

\begin{equation}
\frac{2\nu^{2}}{\nu + 1}\dot{\phi}\ddot{\phi} + (\frac{2\nu}{ \nu + 1} - \frac{\nu^{2}}{(\nu + 1)^{2} })\dot{\nu}\dot{\phi}^{2}+  V'(\phi)\dot{\phi} + 3H \nu \dot{\phi}^{2} = - Q
\end{equation}
where prime denotes differentiation with respect to $ \phi $.
Equations (4), (6) and (9) governs the dynamics of DBI dark energy scalar field $ \phi $, intercating with dark matter.
\section{ Autonomous System}
As the evolution equations are very  complicated in form, so we shall restrict ourselves to study the cosmological evolution through qualitative analysis. For  the phase space  analysis, we introduce   the dimensionless variables x and y [30, 31]

\begin{equation}
x \equiv \frac{\nu\dot{\phi}}{\sqrt{3( 1+\nu)H}} ~~~~~and ~~~~~ y \equiv \frac{\sqrt{V(\phi)}}{\sqrt{3}H}
\end{equation}

Note that  the first Friedman equation in (5) shows the interrelation between the new variables  x and y as
\begin{equation}
x^{2}+ y^{2} +\Omega_{m} = 1
\end{equation}
Using these new variables the above evolution equations can be written as following autonomous system \\
 $~~~~~~~~~~~~~~~~~~~~~~ ~~~~~~~\frac{dx}{dN} = -\lambda y^{2} - \frac{3}{2x}[ \alpha_{m}(1 -x^{2}- y^{2}) +
 \alpha_{\phi}(x^{2}+ y^{2})]\\~~~~~~~~~~~~~~~~~~~~~~~~~~~~~~~~~~~~~~~~ -  \frac{3x}{2} [ \nu_{0}( 1 - x^{2}) + y ^{2}- \omega_{m}(1 -x^{2}- y^{2})]$
 \begin{equation}~~~~~~
 \frac{dy}{dN}= \lambda xy + \frac{3y}{2}[ 1 +\nu_{0}x ^{2} - y ^{2}+ \omega_{m}( 1 -x^{2}- y^{2})]
\end{equation}
$~~~~~~~~~~~~~~~~~~~~~and ~~~~~~\frac{d\nu}{dN}=  2 (\nu - 1 )[\mu - \frac{3}{2x^{2}}[{{( 1 + \nu_{0})x^{2}+ \alpha_{m}( 1 -x^{2}- y^{2})+ \alpha_{\phi}(x^{2}+ y^{2})}}]]$\\

where $  \lambda  = \frac{\sqrt{3 ( 1 + \nu)}}{2\nu}\frac{V'(\phi)}{V(\phi)}~~~~(12a) ~~~~~and ~~~~ \mu = [x\frac{f'(\phi)}{f(\phi)} - \frac{y^{2}}{x}\frac{V'(\phi)}{V(\phi)}]\frac{\sqrt{3 ( 1 + \nu)}}{2\nu}~~~(12b)$

are assumed to be constant. $ N = ln ~a ~~ i.e ~ \frac{d}{dN} = \frac{1}{H}\frac{d}{dt}$ and $ \nu_{0} = \frac{1}{\nu}$.\\
    The critical points for the autonomous system (12) can be obtained by solving $ \frac{dx}{dN}= 0 $ , $ \frac{dy}{dN}= 0 $ and $ \frac{d\nu}{dN}= 0 $ for $\nu \neq 1 $.  However , due to complicated form of the algebraic equations, it is not possible to have any analytic form of the critical points.  Figure 1 depicts the phase space of the autonomous system (12) for some specific values of $  \lambda ,~~ \alpha_{m} , ~\alpha_{\phi} , ~\omega_{m} ,~ \mu $ .  From the figure 1 we see that  most of the trajectories moving towards a point close to $ \nu = 1$ .
\begin{figure}
 \includegraphics[width = 0.6500\linewidth]{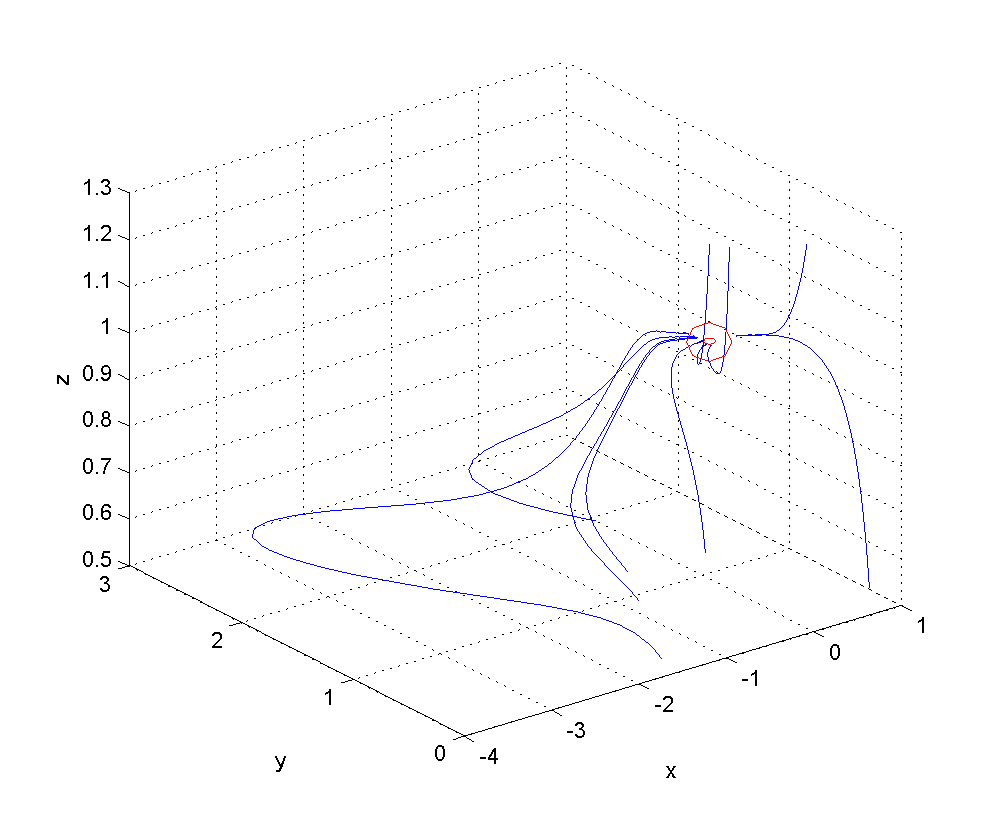}
\caption{\label{}phase space for the autonomous system (12) for $  \lambda = -.95,~~ \alpha_{m} = .001, ~~\alpha_{\phi} = .002, ~~\omega_{m} = 0.77 ~,\mu = .56  $}
  \end{figure}

In particular, choosing  $ \nu = 1$, we have $ \frac{d\nu}{dN} = 0$ and $ \nu_{0} = 1 $.
the system of equations become\\
$~~~~~~~~~~~~~~~~~~~~~~~~~~~~~~~~ \frac{dx}{dN} = -\lambda y^{2} - \frac{3}{2x}[ \alpha_{m}(1 -x^{2}- y^{2}) +
 \alpha_{\phi}(x^{2}+ y^{2})]\\~~~~~~~~~~~~~~~~~~~~~~~~~~~~~~~~~~~~~~~~ -  \frac{3x}{2} [  1 - x^{2} + y ^{2}- \omega_{m}(1 -x^{2}- y^{2})]$
 \begin{equation}~~~~~~
 \frac{dy}{dN}= \lambda xy + \frac{3y}{2}[ 1 + x ^{2} - y ^{2}+ \omega_{m}( 1 -x^{2}- y^{2})]
\end{equation}

  The above first order system of non-linear differential equations (13) can be considered as a 2D autonomous system. In the following section we shall study the autonomous system with some specific choice of the potential $ V(\phi)$.  \\
The equation of state for the  DBI dark energy is  given by
\begin{equation}
   \emph{} \omega_{\phi} = \frac{p_{\phi}}{\rho_{\phi}} = \frac{x^{2}-y^{2}}{x^{2}+ y^{2}}
\end{equation}
and the  effective equation of state ( $\omega_{eff}$) for DBI scalar field plus dark matter  has the expression
\begin{equation}
   \omega_{eff} = \frac{p_{\phi}+ p_{m}}{\rho_{\phi} + \rho_{m}} = x^{2}- y^{2} + \omega_{m}( 1 - x^{2}-y^{2}),
\end{equation}
 with  $\omega_{eff} < -\frac{1}{3}$ for cosmic acceleration .

It should be noted that the physical region in the phase plane is constrained by the requirement that the energy density be non-negative i.e $\Omega_{m}\geq 0$. So  from equation (11),  x and y are restricted to the circular region $ x^{2}+ y^{2}\leq 1$.  The equality sign indicates that there is no longer any dark matter. Further, geometrically equation (11) represents a paraboloid in $ ( \Omega_{m}, x ,y)$-state space and it is possible to divide the 3D-state space into the following invariant sets:\\
  A : $ \Omega_{m} > 0 ~~~~ and~~~~ ~~~~~~ x^{2}+ y^{2} < 1, $ non-vacuum 2D \\
 $ ~~B :  \Omega_{m} = 0 ~~~~ and ~~~~~~~~~~~x^{2}+ y^{2} = 1, $ no fluid matter 1D
\section{ Analysis of Critical points  }

To find the critical points of the system (13) we set
\begin{equation}
\frac{dx}{dN} = 0 ~~~and~~~  \frac{dy}{dN}= 0
\end{equation}

As $ \nu $ is chosen to be unity so from equation (12a), the potential  $ V (\phi) $ can be determined as
\begin{equation}
 V(\phi) = V_{0}exp[ \sqrt{\frac{2}{3}}\lambda\phi]~~~~
 \end{equation}

with $ V_{0}$, the integration constant. Using (16) , we have from (13), two non linear  algebraic equations
\begin{equation}
\frac{2}{3}\lambda xy^{2} +  [ \alpha_{m}(1 -x^{2}- y^{2}) +
 \alpha_{\phi}(x^{2}+ y^{2})]+ x^{2} [  1 - x^{2} + y ^{2}- \omega_{m}(1 -x^{2}- y^{2})] = 0
\end{equation}
 \begin{equation}
 y [\frac{2}{3}\lambda x +  1 + x ^{2} - y ^{2}+ \omega_{m}( 1 -x^{2}- y^{2})] = 0
\end{equation}

\begin{table}
\caption{\emph{Equilibrium points, their nature and values of physical parameters  for $  \lambda = -1,~~ \alpha_{m} = .001, ~~\alpha_{\phi} = .002, ~~\omega_{m} = 0.27 ~ $}}

 \begin{tabular}{|c|c|c|c|c|c|}
 \hline
 Equilibrium point & x & y & $  \Omega_{m} $ & $ \omega_{eff}$   & Nature\\\hline
 $ E_{1}$ & -1 & 0 & 0 & 1 &  unstable node  \\\hline
 $ E_{2}$ & 0 & 0 & 1 & .27 & saddle point\\\hline
 $ E_{3}$ & 1 & 0 & 0 & 1 & unstable node \\\hline
 $E_{4}$ & 0.34 & 0.94 & 0 & -0.77   & stable node\\\hline
 $E_{5}$ & 0.34 & -0.94 & 0 & -0.77 & stable node\\\hline

 \end{tabular}
 \end{table}

\begin{table}
\caption{\emph{Equilibrium points, their nature and values of physical parameters  for $  \lambda = -1,~~ \alpha_{m} = .001, ~~\alpha_{\phi} = .002, ~~\omega_{m} = 0.57 , 0.77 ~ $}}

 \begin{tabular}{|c|c|c|c|c|c|}
 \hline
 Equilibrium point & x & y & $  \Omega_{m} $ & $ \omega_{eff}$   & Nature\\\hline
 $ E_{1}$ & -1 & 0 & 0 & 1 &  unstable node  \\\hline
 $ E_{2}$ & 0 & 0 & 1 & 0.57 (0.77 i.e $ \omega_{m}$) & saddle point\\\hline
 $ E_{3}$ & 1 & 0 & 0 & 1 & unstable node \\\hline
 $E_{6}$ & 0.32 & 0.94 & 0.014 & -0.77   & stable node\\\hline
 $E_{7}$ & 0.32 & -0.94 & 0.014 & -0.77 & stable node\\\hline

 \end{tabular}
 \end{table}

\begin{figure}
\begin{minipage}{.45\textwidth}
 \includegraphics[width = 1.100\linewidth]{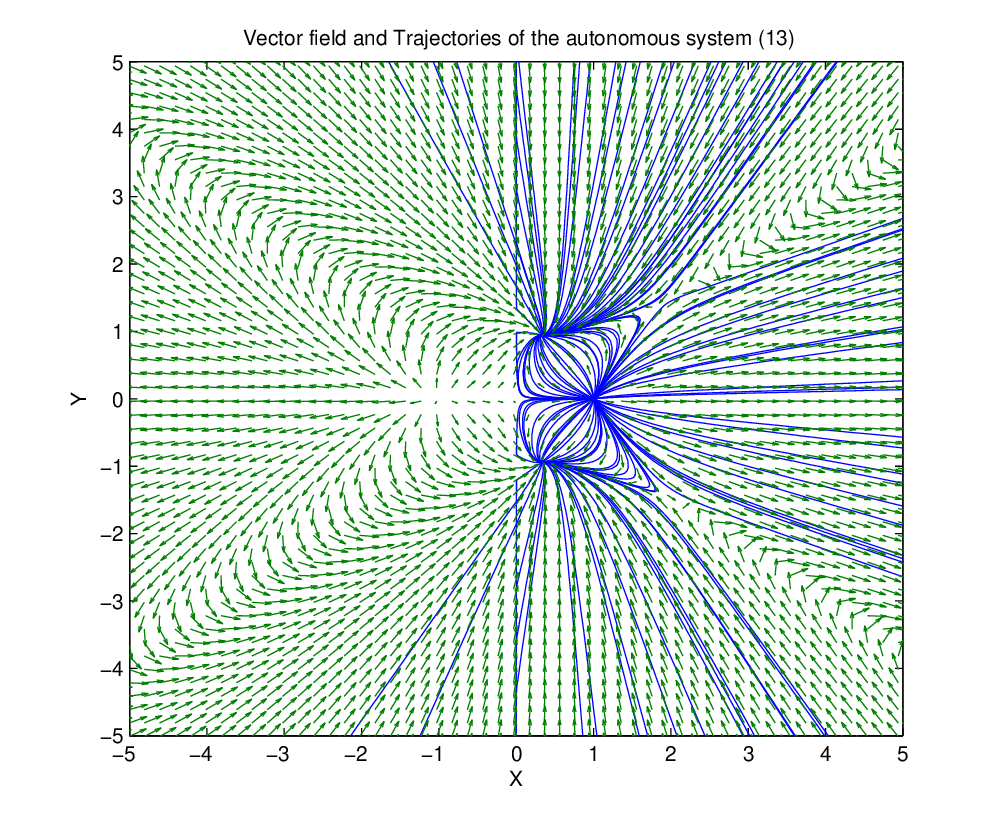}
\caption{\label{}  phase portrait for  $ \lambda =-1,~ \alpha_{m} =.001, ~\alpha_{\phi} = .002,~
                   \omega_{m} = 0.27  $    }
 \end{minipage}
 \begin{minipage}{.45\textwidth}
 \includegraphics[width = 1.10\linewidth]{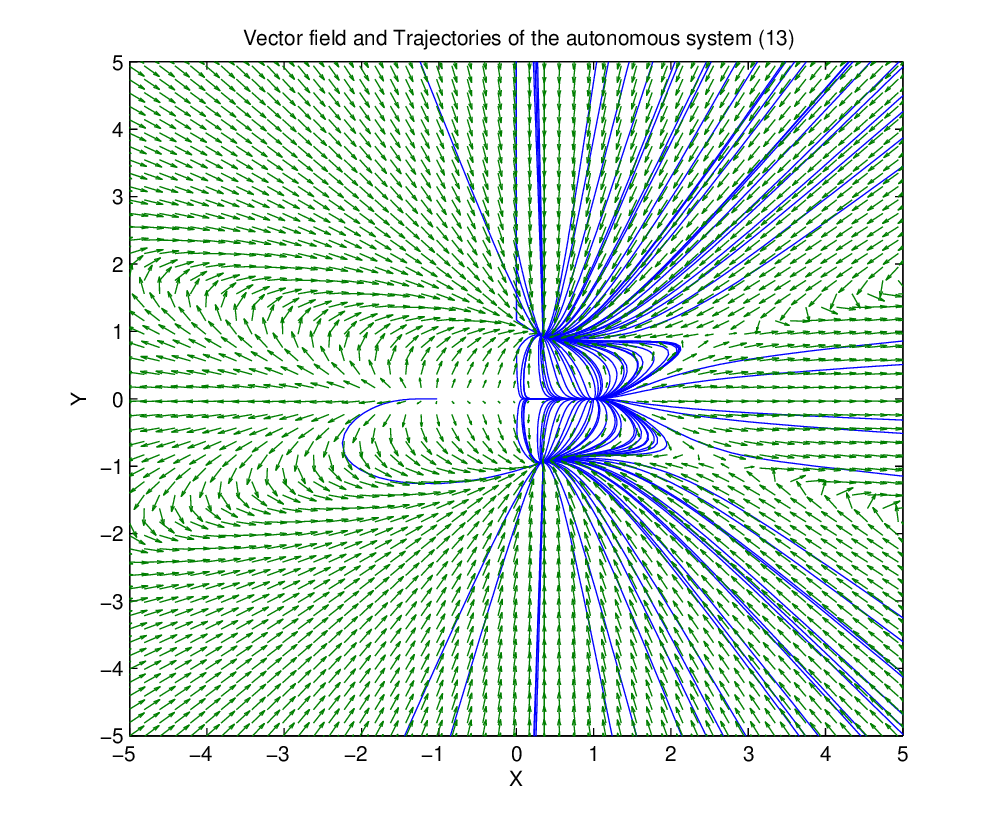}
\caption{\label{}  phase portrait for $ \lambda = -1, ~\alpha_{m} = .001, ~\alpha_{\phi} = .002, ~\omega_{m} = .77  $      }
\end{minipage}

 \end{figure}

 \begin{table}
\caption{\emph{Equilibrium points, their nature and values of physical parameters  for $  \lambda = -2,~~ \alpha_{m} = .001, ~~\alpha_{\phi} = .002, ~~\omega_{m} = 0.57 ~ $}}

 \begin{tabular}{|c|c|c|c|c|c|}
 \hline
 Equilibrium point & x & y & $  \Omega_{m} $ & $ \omega_{eff}$   & Nature\\\hline
 $ E_{1}$ & -1 & 0 & 0 & 1 &  unstable node  \\\hline
 $ E_{2}$ & 0 & 0 & 1 & .57 & saddle point\\\hline
 $ E_{3}$ & 1 & 0 & 0 & 1 & unstable node \\\hline
 $E_{8}$ & 0.66 & 0.75 & 0.002 & -0.125   & stable node\\\hline
 $E_{9}$ & 0.66 & -0.75 & 0.002 & -0.125 & stable node\\\hline

 \end{tabular}
 \end{table}

  \begin{table}
\caption{\emph{Equilibrium points, their nature and values of physical parameters  for $  \lambda = -2,~~ \alpha_{m} = .001, ~~\alpha_{\phi} = .002, ~~\omega_{m} = 0.77 ~ $}}

 \begin{tabular}{|c|c|c|c|c|c|}
 \hline
 Equilibrium point & x & y & $  \Omega_{m} $ & $ \omega_{eff}$   & Nature\\\hline
 $ E_{1}$ & -1 & 0 & 0 & 1 &  unstable node  \\\hline
 $ E_{2}$ & 0 & 0 & 1 & .77 & saddle point\\\hline
 $ E_{3}$ & 1 & 0 & 0 & 1 & unstable node \\\hline
 $E_{8}$ & 0.65 & 0.75 & 0.002 & -0.138   & stable node\\\hline
 $E_{9}$ & 0.65 & -0.74 & 0.002 & -0.138 & stable node\\\hline

 \end{tabular}
 \end{table}

 \begin{figure}
\begin{minipage}{.45\textwidth}
 \includegraphics[width = 1.100\linewidth]{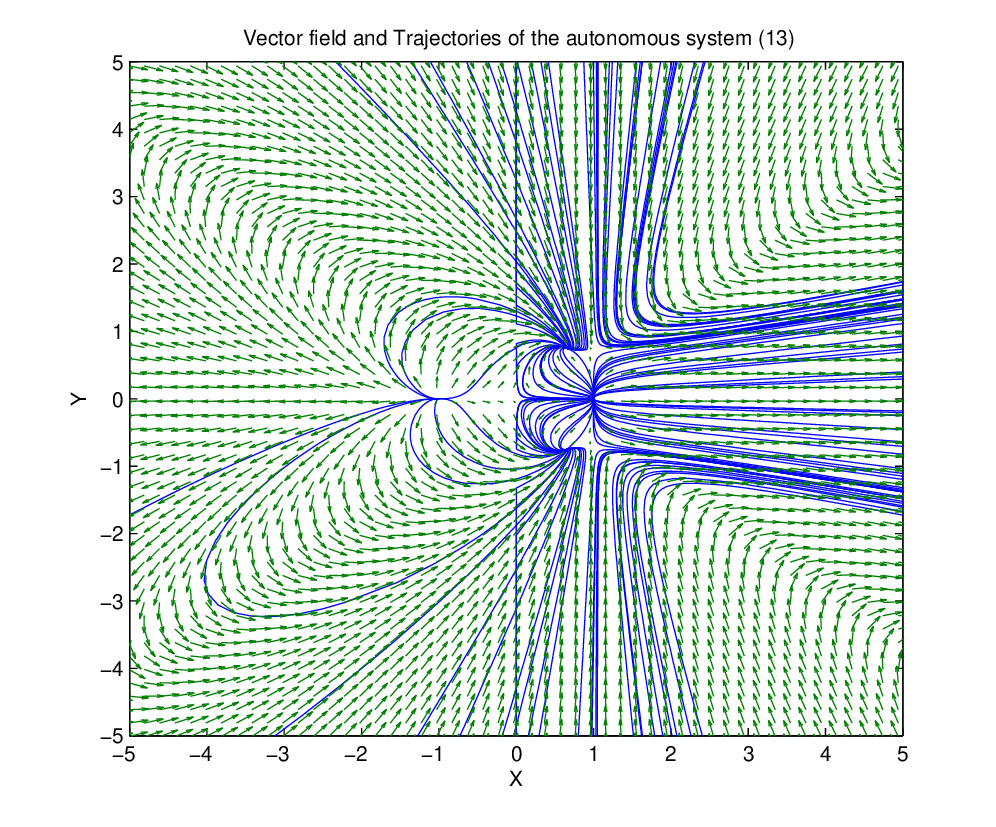}
\caption{\label{} phase portrait for  $ \lambda =-2, ~\alpha_{m} =.001,~ \alpha_{\phi} = .002,~
                   \omega_{m} = 0.27  $ }
 \end{minipage}
 \begin{minipage}{.45\textwidth}
 \includegraphics[width = 1.10\linewidth]{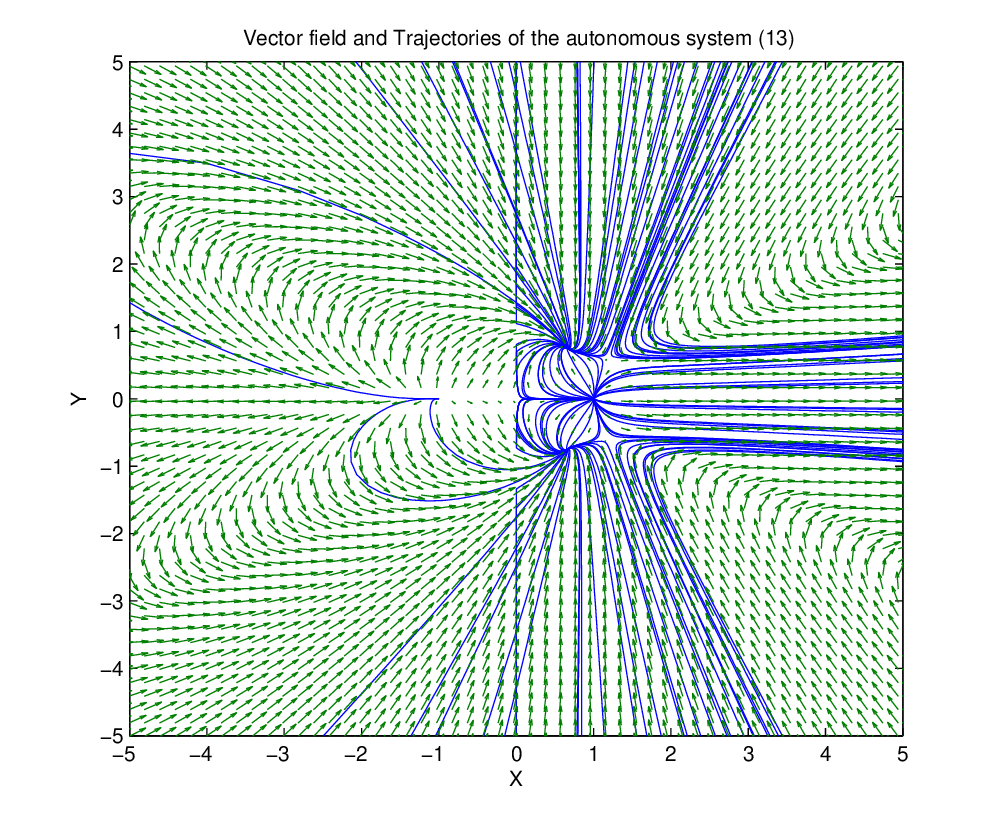}
\caption{\label{} phase portrait for  $ \lambda =-2, ~\alpha_{m} =.001, ~\alpha_{\phi} = .002,~
                   \omega_{m} = 0.57  $   }
\end{minipage}

 \end{figure}


We instantly see that (0,0) is a critical point [32, 33, 34].
For y = 0, we have $ x = \pm ~~x_{0} $ with $ x_{0} \simeq \sqrt{ 1 + \frac{\alpha_{\phi}}{1- \omega_{m}}}$. In deriving $ x_{0} $, we have neglected the product $\alpha_{\phi}.\alpha_{m} $ as $\alpha_{\phi}$, $ \alpha_{m} \ll 1$ . 

For non zero y, the critical points $ (x_{c}, y _{c})$ are obtained as follows: $ y_{c}$ is related to $ x_{c}$ by the quadratic relation
\begin{equation}
 y_{c}^{2 }=  1 + \frac{1}{ 1 + \omega_{m}}[ \frac{2}{3}\lambda x_{c} + ( 1 - \omega_{m})x_{c} ^{2}]
\end{equation}
and $ x_{c}$ satisfies the cubic equation
\begin{equation}
\lambda x^{3} +  \frac{3}{2}x^{2}[\frac{2\lambda^{2}}{9}+ \mu_{0}^{2}] + \frac{1}{2}\lambda x \mu_{0}^{2}+\frac{3}{4}
 \alpha_{\phi}(1 + \omega_{m}) = 0
\end{equation}
 with $ \mu_{0}^{2} =  1 + \alpha_{\phi} + \omega_{m}-  \alpha_{m}$ .

 Note that if $ \lambda > 0$, then the above cubic equation in x has no real positive root , but it may have three or one negative real root. On the other hand , if $ \lambda < 0 $, then x may have three or one positive real root but can not have any negative real root.
 \begin{table}
\caption{\emph{Equilibrium points, their nature and values of physical parameters  for $  \lambda = -3,~~ \alpha_{m} = .001, ~~\alpha_{\phi} = .002, ~~\omega_{m} = 0.27 ~ $}}

 \begin{tabular}{|c|c|c|c|c|c|}
  \hline
 Equilibrium point & x & y & $  \Omega_{m} $ & $ \omega_{eff}$   & Nature\\\hline
 $ E_{1}$ & -1 & 0 & 0 & 1 &  unstable node  \\\hline
 $ E_{2}$ & 0 & 0 & 1 & .27 & saddle point\\\hline
 $ E_{3}$ & 1 & 0 & 0 & 1 & unstable node \\\hline
 $E_{10}$ & 0.62 & 0.49 & .375 & 0.245   & stable spiral\\\hline
 $E_{11}$ & 0.62 & -0.49 & 0.375 & 0.245 & stable spiral\\\hline

 \end{tabular}
 \end{table}

\begin{table}
\caption{\emph{Equilibrium points, their nature and values of physical parameters  for $  \lambda = -3,~~ \alpha_{m} = .001, ~~\alpha_{\phi} = .002, ~~\omega_{m} = 0.57 ~ $}}

 \begin{tabular}{|c|c|c|c|c|c|}
 \hline
 Equilibrium point & x & y & $  \Omega_{m} $ & $ \omega_{eff}$   & Nature\\\hline
 $ E_{1}$ & -1 & 0 & 0 & 1 &  unstable node  \\\hline
 $ E_{2}$ & 0 & 0 & 1 & .57 & saddle point\\\hline
 $ E_{3}$ & 1 & 0 & 0 & 1 & unstable node \\\hline
 $E_{12}$ & 0.75 & 0.44 & 0.244 & 0.51   & stable spiral\\\hline
 $E_{13}$ & 0.75 & -0.44 & 0.244 & 0.51 & stable spiral\\\hline

 \end{tabular}
 \end{table}

  \begin{figure}
\begin{minipage}{.45\textwidth}
 \includegraphics[width = 1.000\linewidth]{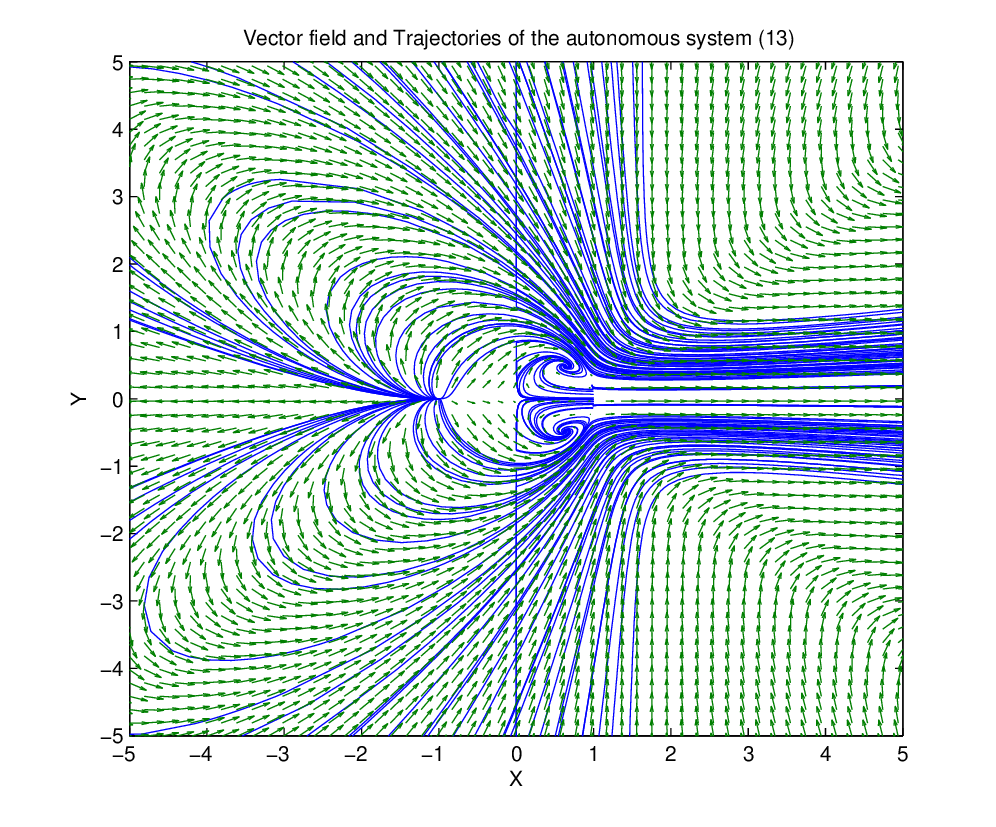}
\caption{\label{} phase portrait for  $ \lambda =-3,~ \alpha_{m} =.001, ~\alpha_{\phi} = .002,
                  ~ \omega_{m} = 0.27  $ }
 \end{minipage}
\begin{minipage}{.45\textwidth}
 \includegraphics[width = 1.000\linewidth]{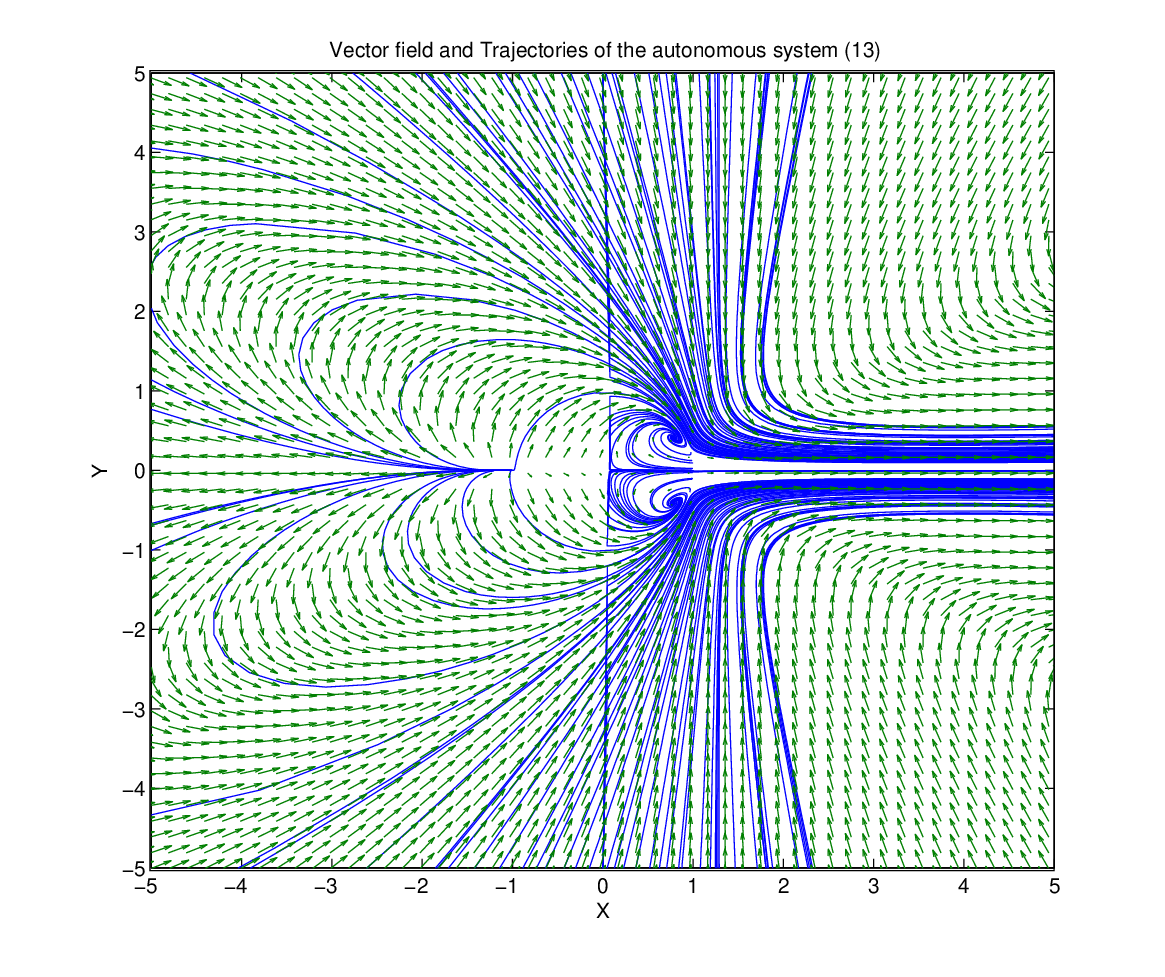}
\caption{\label{} phase portrait for  $ \lambda =-3,~
  \alpha_{m} =.001, ~\alpha_{\phi} = .002,~
                   \omega_{m} = 0.57  $ }
 \end{minipage}
 \end{figure}

\begin{table}
\caption{\emph{Equilibrium points, their nature and values of physical parameters  for $  \lambda = 1,~~ \alpha_{m} = .001, ~~\alpha_{\phi} = .002, ~~\omega_{m} = 0.27 ~ $}}

 \begin{tabular}{|c|c|c|c|c|c|}
 \hline
 Equilibrium point & x & y & $  \Omega_{m} $ & $ \omega_{eff}$   & Nature\\\hline
 $ E_{1}$ & -1 & 0 & 0 & 1 &  unstable node  \\\hline
 $ E_{2}$ & 0 & 0 & 1 & .27 & saddle point\\\hline
 $ E_{3}$ & 1 & 0 & 0 & 1 & unstable node \\\hline
 $E_{14}$ & -0.34 & 0.94 & 0 & -0.77   & stable node\\\hline
 $E_{15}$ & -0.34 & -0.94 & 0 & -0.77 & stable node\\\hline

 \end{tabular}
 \end{table}

  Phase portrait for the system is drawn for different values of $ \lambda , \alpha_{\phi}, \alpha_{m} $ and  $ \omega_{m} $ [ Fig 2 to Fig 7].  The Critical points  for the system (13), corresponding   density parameter for dark matter, effective equation of state parameter , stability are presented in tables I to VII . We have checked as expected that   there is no significant change in critical points or in their stability for variation of $\alpha_{\phi} $ or $\alpha_{m} $ , but critical points and  stability may vary if $ \lambda $ and $ \omega_{m} $ change.  If we keep $ \lambda $ fixed and vary $  \omega_{m} $ no change or insignificant change in critical points is noticed. We may conclude that interaction has no significant effect on the existence as well as on the stability of the critical points.
 We should mention here that since the system is complicated , so  the analysis is done using  some numerical values of the parameters involved and  by the phase space  only.

   For different values of $ \lambda~, \omega_{m} $ , $ E_{1} $  and $ E_{3} $  are either unstable node or saddle point. Tables show that  critical point $ E_{2}$ corresponding to $ \Omega_{m}= 1 $ is saddle in nature. So in the absence of dark energy i.e when the universe is  dominated by dark matter , the present DBI model can not describe the evolution of the universe.
   The critical points $ E_{4}, E_{5}$ in table I, or equivalently  $ E_{6},E_{7}$ in table II and $ E_{14}, E_{15}$ in table VII   are interesting in present cosmological scenario. These two critical points represent late time accelerated expansion of the universe and are stable in nature being a stable node.  Critical points $E_{4}, E_{5}$  or $ E_{14},E_{15}$  are completely characterized by DE (the DBI scalar field). It should be noted that these critical points describe cosmic evolution only in quintessence era. Critical points $  E_{8}$ , $ E_{9}$ , $ E_{10}$, $ E_{11}$, $ E_{12}$, $ E_{13}$ are stable in nature but do not correspond to late time acceleration. 
    Further, it may be noted that there does not exist any critical point corresponding to inflationary scenario in early universe. \\
    On the other hand, to examine whether the present model can be extended to the phantom barrier  (i.e $\omega_{eff} \simeq -1 $), we see from equation (15) that it is possible if $ x = 0, y = \pm 1 $ .  Also the points $ (0, \pm 1 )$ will satisfy  the non-linear algebraic equations (18) and (19) provided the coupling parameter $ \alpha_{\phi} $ in the interaction term (8) is zero . The phase portrait corresponding these two critical points are shown in figure 8 and 9. Nature of these two critical points and  corresponding physical parameters   are  given in tabular form in table VIII .

    \begin{table}
\caption{\emph{Equilibrium points, their nature  for $  ~~\alpha_{\phi} = 0, $}}
 \begin{tabular}{|c|c|c|c|c|c|}
 \hline
 Equilibrium point & x & y & $  \Omega_{m} $ & $ \omega_{eff}$   & Nature\\\hline

 $E_{16}$ &~ 0 & ~1~ & 0 ~& -1   & stable node\\\hline
 $E_{17}$ &~ 0 & -1~ & 0 ~& -1 & stable node\\\hline

 \end{tabular}
 \end{table}

  \begin{figure}
\begin{minipage}{.45\textwidth}
 \includegraphics[width = 1.000\linewidth]{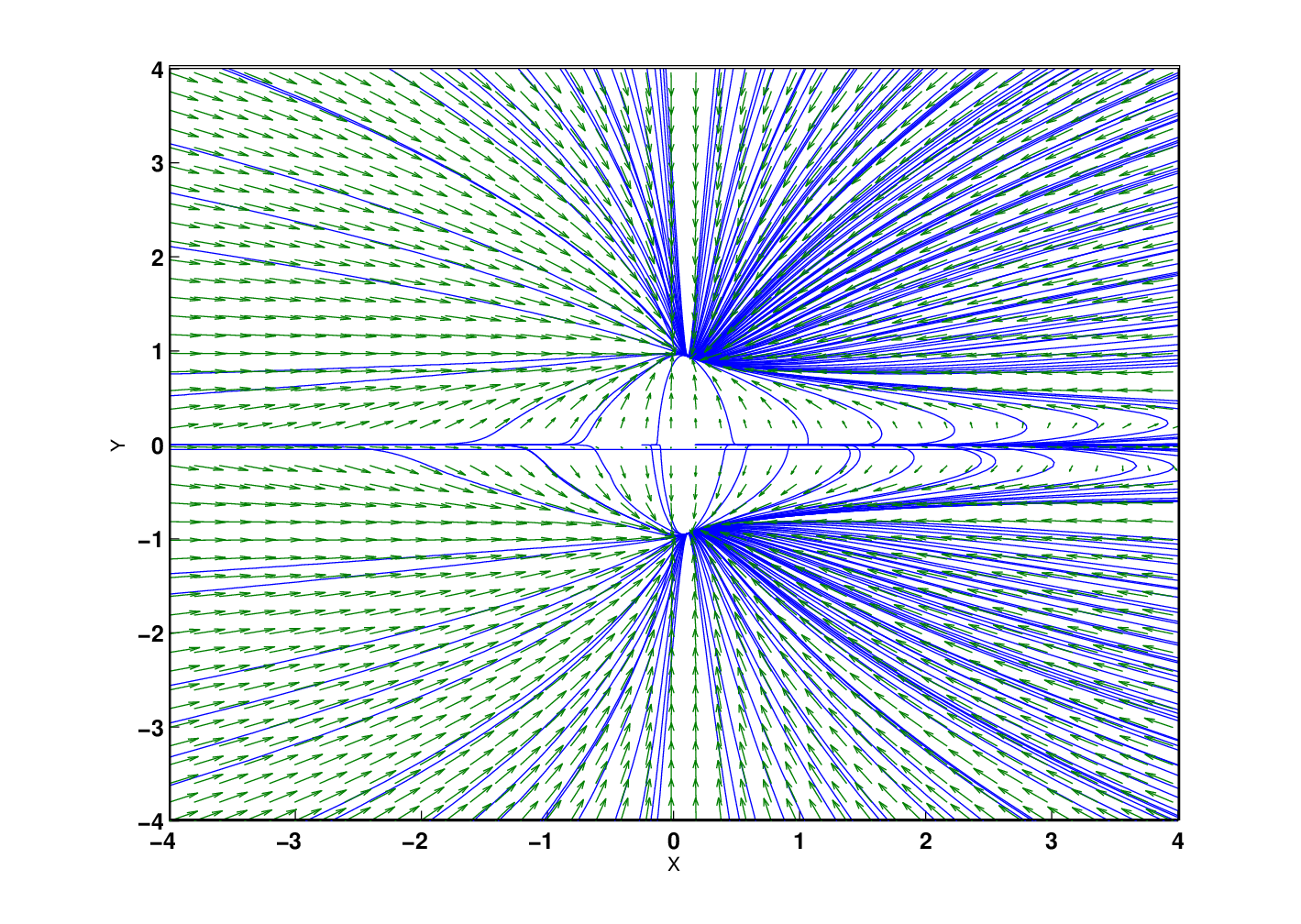}
\caption{\label{} phase portrait for  $ \lambda =-1,~ \alpha_{m} =.001, ~\alpha_{\phi} = 0,
                  ~ \omega_{m} = 0.97  $ }
 \end{minipage}
\begin{minipage}{.45\textwidth}
 \includegraphics[width = 1.000\linewidth]{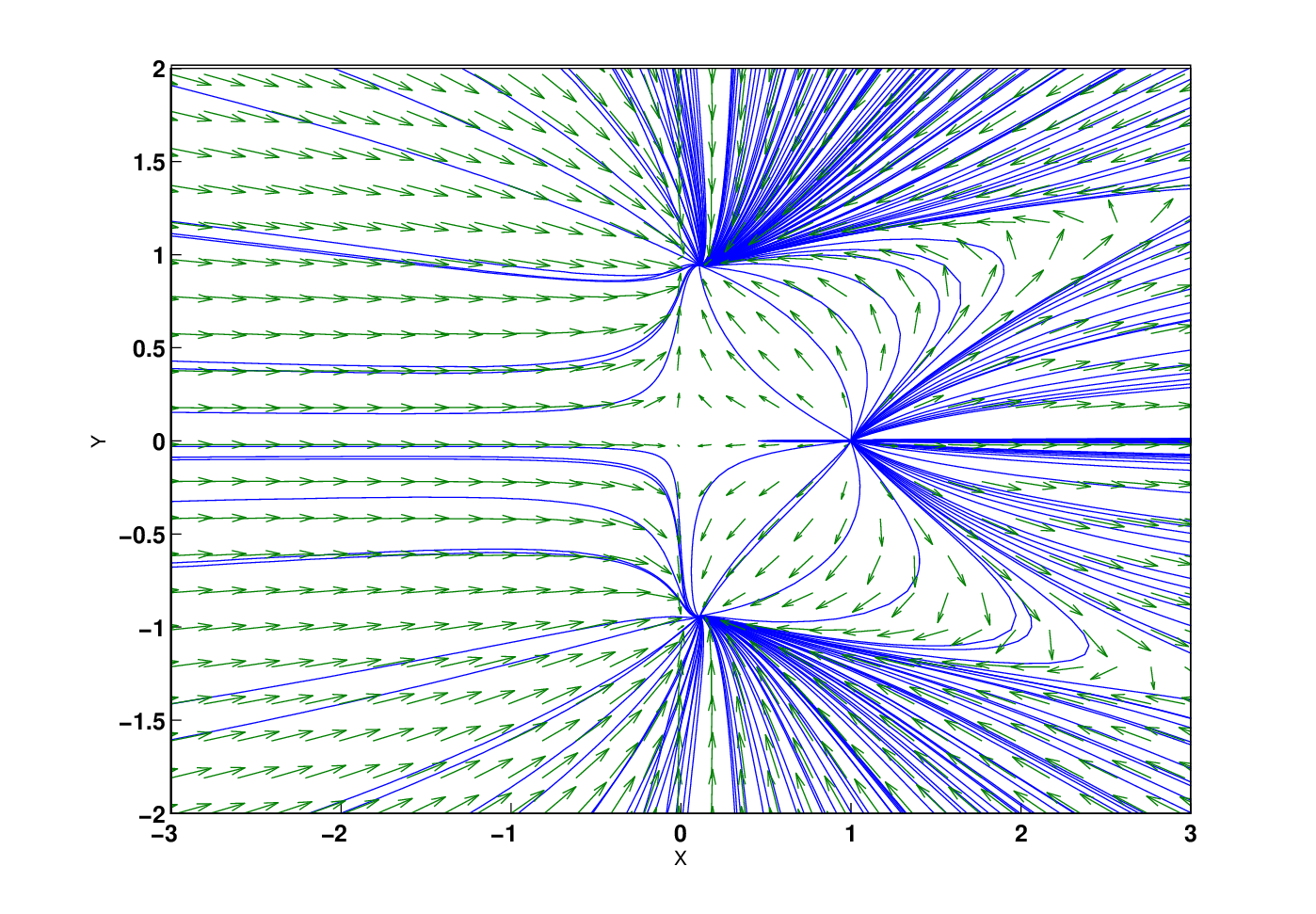}
\caption{\label{} phase portrait for  $ \lambda =-1,~  \alpha_{m} =.001, ~\alpha_{\phi} = 0,~
                   \omega_{m} = 0.27  $ }
 \end{minipage}
 \end{figure}

 Thus both the critical points are stable in nature and correspond to $ \Lambda CDM$ model (phantom barrier).

    Therefore, We may conclude that  the present interacting DBI model can explain late time acceleration only upto phantom barrier but not in phantom domain which is the possible region for DE by recent observations.

 %

\begin{acknowledgments}
The authors are thankful to IUCAA, Pune for warm hospitality and facilities at the library as major part of the work is done during a visit to IUCAA. The authors are also thankful to UGC-DRS programme, Department of Mathematics, Jadavpur University.
\end{acknowledgments}
\section{References }

\begin{thebibliography}{}
\bibitem{RefA}
A. G. Riess et al., Astron. J. 116, 1009 (1998)
\bibitem{RefB}
S. J. Perlmutter et al., Astrophys. J. 517, 565 (1999);
\bibitem{RefA}
 D. N. Spergel et al.(WMAP Collaboration), Astrophys. J. Suppl. Ser. 170, 377 (2007);
\bibitem{RefB}
E. Komatsu et al (WMAP Collaboration), Astrophys. J. Suppl. Ser 180, 330 (2009)
\bibitem{RefA}
 Pereival et al , Mon. Not. R. Astr. S. 381, 1053 (2007)
\bibitem{RefC}
S. Weinberg, Rev. Mod. Phys. 61, 1 (1989)
 \bibitem{RefB}
V. Sahni and A. Starobinsky, Int. J. Mod. Phys. D 9, 377 (2000)
\bibitem{RefC}
 S. M. Carroll, Liv. Rev. Lett. 4, 1 (2001)
 \bibitem{RefI}
 B. Ratra and P. J. E. Peebles  Phys. Rev. D 37,  3406  (1988)

\bibitem{RefC}
 R. R. Caldwell, R. Dave and P. J. Steinherdt, Phys. Rev. Lett 80, 1582 (1998)
 \bibitem{RefC}
I. Zlatev, L. Wang and P. J. Steinhardt, Phys.  Rev. Lett. 82, 896 (1999)
\bibitem{RefD}
C. Armendariz-Picon, V. Mukhanov and P. J. Steinhardt, Phys. Rev. Lett. 85, 4438 (2000)
\bibitem{RefD}
C. Armendariz-Picon, V. Mukhanov and P. J. Steinhardt, Phys.  Rev. D 63, 103510 (2001)
 \bibitem{RefM}
R. R. Caldwell, Phys. Lett. B 545, 23 (2002)
\bibitem{RefN}
R. R. Caldwell, M.Kamionkowski and N. N. Weinberg, Phys. Rev. Lett. 91, 071301 (2003)
\bibitem{RefD}
A. Sen, JHEP 0207, 065(2002)
\bibitem{RefD}
 T. Padmanabhan, Phys. Rev. D 66, 021301 (2002)
 \bibitem{RefB}
F. Piazza and S. Tsujikawa, JCAP 0407, 004 (2004)
\bibitem{RefE}
E. Elizalde, S. Nojiri and S. D. Odintsov, Phys. Rev. D 70, 043539( 2004)
\bibitem{RefE}
 S. Nojiri and S. D. Odintsov and S. Tsujikawa, Phys. Rev. D 71, 063004(2005)
\bibitem{RefL}
L. Amendola and S. Tsujikawa, Dark Energy: Theory and Observations (Cambridge Univ. Press, Cambridge, England, 2010)
\bibitem{RefD}
E. Silverstein and D. Tong, Phys. Rev. D 70, 103505 (2004)
\bibitem{RefD}
X. Chen, Phys. Rev. D 71, 063506 (2005)
\bibitem{RefB}
Luis P. Chimento, Ruth Lazkoz, and Martín G. Richarte, Phys. Rev. D 83, 063505 (2011)

\bibitem{RefD}
C. Armendariz-Picon, T. Damour and  V. Mukhanov , Phys.  Lett.  B 458, 209 (1999)

\bibitem{RefC}
E. Guendelman, D. Singleton and N. Yongram, JCAP 11, 044 (2012)
\bibitem{RefC}

G. Olivares, F.  Atrio-Barandela and D. Pavon , Phys.Rev. D 71, 063523 (2005)

\bibitem{RefC}
G. Olivares, F.  Atrio-Barandela and D. Pavon , Phys.Rev. D 74, 043521 (2006)


\bibitem{RefC}
D.  Pavon, and B. Wang,  Gen. Rel. Grav. 41, 1(2009)

\bibitem{RefC}
E. J. Copeland,  A. R. Liddle  and  D. Wands,  Phys. Rev. D 57, 4686 (1998).

\bibitem{RefC}
C. Kaeonikhom, D. Singleton, S.  V. Sushkov and N. Yongram, Phys. Rev. D 86, 124049 (2012)

\bibitem{RefB}
M. Hirsch and S. Smale, " Differential Equations, Dynamical Systems and Linear Algebra " (1974)(2nd edn, New york: Academic)

\bibitem{RefB}
D. K. Arrowsmith and C. M. Place, " An Introduction  to Dynamical Syatems" (1990)( Cambridge Univ. Press, Cambridge)
\bibitem{RefB}
S. Wiggins, "Introduction to  Applied Nonlinear Dynamical Systems and Chaos" (2003)( 2nd Edn. Berlin; Springer)


\end{thebibliography}

\end{document}